\documentclass[12pt]{article}
\usepackage{cite}
\newcommand{\xrm}[1]{{\textstyle \mbox{\rm #1}}}
\newcommand{\fnd}[2]{\frac{\textstyle #1}{\textstyle #2}}
\begin{document} \baselineskip .7cm
\title{\bf A low-lying scalar meson nonet in a unitarized meson model}

\author{
E.~van Beveren$^{1,2}$, T.~A.~Rijken, K.~Metzger$^{3}$, C.~Dullemond,\\
{\normalsize\it Institute for Theoretical Physics, University of Nijmegen}\\
{\normalsize\it NL-6525 ED, Nijmegen, The Netherlands}\\ [.3cm]
G.~Rupp$^{4,5}$,\\
{\normalsize\it Zentrum f\"{u}r Interdisziplin\"{a}re Forschung,
Universit\"{a}t Bielefeld}\\
{\normalsize\it D-4800, Bielefeld, Federal Republic of Germany}\\ [.3cm]
\and
J.~E.~Ribeiro$^{4,6}$\\
{\normalsize\it Centro de F\'{\i}sica da Mat\'{e}ria Condensada}\\
{\normalsize\it Avenida do Prof. Gama Pinto, 2, P-1699, Lisboa, Portugal}
}
\footnotetext[1]{present address:
Centro de F\'{\i}sica Te\'{o}rica,
Departamento de F\'{\i}sica, Universidade de Coimbra,
P-3004-516 Coimbra, Portugal.}
\footnotetext[2]{e-mail: eef@teor.fis.uc.pt}
\footnotetext[3]{present address:
Philips Company, Eindhoven, The Netherlands}
\footnotetext[4]{present address:
Centro de F\'{\i}sica das Interac\c{c}\~{o}es
Fundamentais, Instituto Superior T\'{e}cnico, Universidade T\'{e}cnica de
Lisboa, Edif\'{\i}cio Ci\^{e}ncia, P-1049-001 Lisboa, Portugal.}
\footnotetext[5]{e-mail: george@ist.utl.pt}
\footnotetext[6]{e-mail: EmilioRibeiro@mail.ist.utl.pt}

\date{preprint, 20 August 1985\\ - \\
published in \\ Zeitschrift f\"{u}r Physik {\bf C} - Particles and Fields
{\bf 30}, 615--620 (1986)}
\maketitle

\begin{abstract}
A unitarized nonrelativistic meson model which is successful for the
description of the heavy and light vector and pseudoscalar mesons yields,
in its extension to the scalar mesons but for the same model parameters,
a complete nonet below 1 GeV. In the unitarization scheme, real
and virtual meson-meson decay channels are coupled to the quark-antiquark
confinement channels.
The flavor-dependent harmonic-oscillator confining potential itself has bound
states $\epsilon$(1.3 GeV), $S$(1.5 GeV), $\delta$(1.3 GeV),
$\kappa$(1.4 GeV), similar to the results of other bound-state
$q\bar{q}$ models.
However, the full coupled-channel equations show poles at
$\epsilon$(0.5 GeV), $S$(0.99 GeV), $\delta$(0.97 GeV),
$\kappa$(0.73 GeV). Not only can these pole positions be calculated in
our model, but also cross sections and phase shifts in the meson-scattering
channels, which are in reasonable agreement with the available data for
$\pi\pi$, $\eta\pi$ and $K\pi$ in $S$-wave scattering.
\end{abstract}
\clearpage

\section{Introduction}

The rich structure in meson-meson scattering at intermediate energies has
stimulated many theoreticians to fit the existing quark models to the
experimental results
\cite{NPB121p514,PRL49p624,PLB51p71,PRD19p2678,PLB93p65,PRD15p267,PLB60p201}.
Especially $S$-wave
meson-meson scattering shows structures which are very intriguing
\cite{PLB50p1,PRD7p1279,NPB75p189,NPB79p301,NPB64p134,PRL47p1378,
NPB95p322,NPB144p253,NPB101p333,RMP56pS1,NPB133p490,PRD9p1872}.
Detailed phase-shift analyses reveal two
pronounced scalar mesonic resonances below 1 GeV,
namely the $S$(975) resonance in $\pi\pi$
\cite{PLB50p1,PRD7p1279,NPB75p189,NPB79p301,NPB64p134,NPB95p322}
and the $\delta$(980) in $\eta\pi$ \cite{NPB144p253,NPB101p333}
$S$-wave scattering. The other relevant resonances that appear nowadays in
the tables of particle properties \cite{RMP56pS1} are the $\epsilon$(1300)
in $\pi\pi$ and $\kappa$(l350) in $K\pi$
\cite{NPB133p490,PRD9p1872}
$S$-wave scattering.

It is well known that these particles cause severe problems if one
wants to understand them as quark+antiquark ($q\bar{q}$) states.
For instance, confronted with the $SU(3)_\xrm{{\tiny flavor}}$ quark model,
the resonance positions do not fit the quadratic or linear Gell-Mann--Okubo
mass relations \cite{PRD19p2678} (see, however, \cite{PLB51p71}).
A possible resonance in $\pi\pi$ $S$-wave scattering at 600~MeV,
which was poorly recognized in early analyses \cite{PLB50p1}, disappeared
from the tables of particle properties in the seventies. Nevertheless,
some years later this resonance revived within the bag model, due to a
solution for the scalar-meson problem presented by Jaffe \cite{PRD15p267},
who pointed out that these resonances stem from $qq\bar{q}\bar{q}$ states.
The large binding energy which is assumed for such configurations
makes the low masses required by experiment possible. All kinds
of quark configurations \cite{PRD15p267,PRD21p1370,PRL40p1543}
and gluon-gluon bound states \cite{PLB60p201} might exist, other than the
standard $q\bar{q}$ for mesons and $qqq$ for baryons. This probably no one
doubts, but there is no experimental evidence that they should couple
significantly to hadron-hadron scattering \cite{PRL49p624,PLB93p65}.

To select the $\epsilon$(1300) resonance as the isosinglet
partner of the $S$(975), rather than the $\epsilon$(600), is
probably the result of bag-model interference with the
analysis of the $\pi\pi$ $S$-wave scattering data, because in
the bag model, and also in other bound-state hadron
models, the lowest $J^{PC}=0^{++}$ isospin-zero $q\bar{q}$ object
fits better with a total mass of about 1.3 GeV
\cite{PRD21p1370,PhDAAerts}.

In this paper we will show that we have no
difficulties to explain the scalar mesons within our
unitarized quark model, and to interpret them as $q\bar{q}$
states with a meson-meson admixture. However,
neither the model nor the data exclude poles in
the scattering matrix which do not appear in the
tables but nevertheless might be interpreted as resonances.
\clearpage

\section{The Model}

The unitarized quark model is described in many
articles. We will therefore confine ourselves to only
briefly discuss the main features here and to give a
complete list of references,
\cite{PRD27p1527,PRD21p772,CPC27p377,LNP211p182,LNP211p331,PRD25p2406}.
In our treatment,
meson-meson scattering processes couple to $q\bar{q}$
quark configurations or mesons via the annihilation
and creation of a $q\bar{q}$ pair out of the vacuum. The
reverse coupling describes the decay process of a
meson or the coupling of a meson to its virtual
decay channels. In \cite{PRD27p1527} and
\cite{PRD21p772},
the explicit form of a
multichannel Schr\"{o}dinger description of such a system
is given (see also Appendix~\ref{appendix}).
Several meson-meson scattering channels are, via the
{\it QPC} \/mechanism \cite{NPB10p521}, coupled to
permanently closed $q\bar{q}$ channels with the same quantum numbers.
It is also shown in \cite{PRD27p1527} how to account for relativistic
effects, and for the effects of one-gluon exchange in the $q\bar{q}$
channels.

Scattering matrices, phase shifts, cross sections and wave functions
can be calculated from the Schr\"{o}dinger equation by an approximative
method \cite{CPC27p377,LNP211p182} which leads to an $S$-matrix
that is explicitly analytic in the complex energy plane, and unitary.

Phase shifts and cross sections can be checked to be in good agreement
with the data if available. In other cases, the pole positions of the
scattering matrix can be compared with the bound-state and resonance
positions found by experiment. Wave functions may be compared with
those expected from leptonic decays.

It has been our observation \cite{PRD27p1527} that the properties of the
$J^{PC}=1^{--}$ and $J^{PC}=0^{-+}$ mesonic resonances are reasonably
well described with a few model parameters:
the effective quark masses, where the effective up and down masses
can be taken equal, one universal harmonic-oscillator frequency which
describes the confining force in the permanently closed $q\bar{q}$ channels
for all possible flavor configurations, and two or three parameters
to describe the coupling of the scattering sector to the confinement sector.

In this investigation, we apply the model to $S$-wave meson-meson scattering.
The quark and the antiquark in the permanently closed channel(s)
move in relative $P$-waves, whereas the mesons in the scattering channels
are in relative $S$- and $D$-waves. For the $\epsilon$ and $S$
we use one Schr\"{o}dinger equation with two permanently closed
channels, one for the $n\bar{n}$ pair and one for the $s\bar{s}$ pair.
The mixing occurs in our model quite naturally via the coupling
to scattering channels which contain strange mesons.
We will discuss the results furtheron.

In the first place, we do not alter the effective quark masses,
nor the universal harmonic oscillator frequency.
The only place where we allow some minor changes, if necessary, is in
the potential which couples the confinement and the decay sectors.
For the vector and pseudoscalar mesons, the so-called color splitting
could be accounted for by a component of this potential.
As a result of our calculations for the scalar mesons, we conclude
that in their case other possible interactions seem to compensate
the effects of color splitting. So we decide to set to zero the parameter
which regulated the color splitting in the case of the vector and pseudoscalar
particles, and not to take other contributions into account. The only
legitimations of this procedure are the facts that the results came out
reasonable and that it is not very relevant for the point we want to make in
the present paper. The other two parameters in the coupling potential remain
unaltered with respect to the corresponding parameters in the case of vector
and pseudoscalar mesons.
\clearpage

\section{Results}

\begin{figure}[ht]
\normalsize
\begin{center}
\begin{picture}(283.46,180.08)(-50.00,-30.00)
\put(34.29,-5.52){\makebox(0,0)[tc]{0.4}}
\put(91.45,-5.52){\makebox(0,0)[tc]{0.6}}
\put(148.61,-5.52){\makebox(0,0)[tc]{0.8}}
\put(-5.52,24.24){\makebox(0,0)[rc]{50}}
\put(-5.52,48.48){\makebox(0,0)[rc]{100}}
\put(-5.52,72.73){\makebox(0,0)[rc]{150}}
\put(-5.52,96.97){\makebox(0,0)[rc]{200}}
\put(-5.52,121.21){\makebox(0,0)[rc]{250}}
\put(234.34,-5.52){\makebox(0,0)[tr]{GeV}}
\put(113.16,-20.07){\makebox(0,0)[tc]{$\pi\pi$ invariant mass}}
\put(-5.52,140.59){\makebox(0,0)[tr]{deg}}
\put(5.52,132.57){\makebox(0,0)[tl]{$\pi \pi$ S-wave}}
\tiny
\put(11.43,2.91){\makebox(0,0){$\bullet$}}
\put(22.86,8.73){\makebox(0,0){$\bullet$}}
\put(34.29,5.33){\makebox(0,0){$\bullet$}}
\put(45.72,7.90){\makebox(0,0){$\bullet$}}
\put(57.16,7.76){\makebox(0,0){$\bullet$}}
\put(68.59,10.67){\makebox(0,0){$\bullet$}}
\put(80.02,10.18){\makebox(0,0){$\bullet$}}
\put(91.45,14.55){\makebox(0,0){$\bullet$}}
\put(102.88,17.45){\makebox(0,0){$\bullet$}}
\put(114.31,22.79){\makebox(0,0){$\bullet$}}
\put(125.74,24.73){\makebox(0,0){$\bullet$}}
\put(137.17,47.03){\makebox(0,0){$\bullet$}}
\put(148.61,42.18){\makebox(0,0){$\bullet$}}
\put(160.04,46.06){\makebox(0,0){$\bullet$}}
\put(171.47,45.09){\makebox(0,0){$\bullet$}}
\put(182.90,57.70){\makebox(0,0){$\bullet$}}
\put(194.33,57.70){\makebox(0,0){$\bullet$}}
\put(65.73,21.77){\makebox(0,0){$\circ$}}
\put(71.45,21.38){\makebox(0,0){$\circ$}}
\put(77.16,24.58){\makebox(0,0){$\circ$}}
\put(82.88,24.44){\makebox(0,0){$\circ$}}
\put(88.59,25.65){\makebox(0,0){$\circ$}}
\put(94.31,26.47){\makebox(0,0){$\circ$}}
\put(100.02,27.98){\makebox(0,0){$\circ$}}
\put(105.74,29.24){\makebox(0,0){$\circ$}}
\put(111.45,29.14){\makebox(0,0){$\circ$}}
\put(117.17,32.14){\makebox(0,0){$\circ$}}
\put(122.89,33.31){\makebox(0,0){$\circ$}}
\put(128.60,35.93){\makebox(0,0){$\circ$}}
\put(134.32,39.47){\makebox(0,0){$\circ$}}
\put(145.75,42.91){\makebox(0,0){$\circ$}}
\put(151.46,44.90){\makebox(0,0){$\circ$}}
\put(157.18,47.56){\makebox(0,0){$\circ$}}
\put(162.89,48.14){\makebox(0,0){$\circ$}}
\put(168.61,46.64){\makebox(0,0){$\circ$}}
\put(174.33,50.71){\makebox(0,0){$\circ$}}
\put(180.04,52.51){\makebox(0,0){$\circ$}}
\put(92.88,27.15){\makebox(0,0){$\ast$}}
\put(98.59,29.05){\makebox(0,0){$\ast$}}
\put(104.31,32.04){\makebox(0,0){$\ast$}}
\put(110.03,30.62){\makebox(0,0){$\ast$}}
\put(115.74,33.15){\makebox(0,0){$\ast$}}
\put(121.46,36.30){\makebox(0,0){$\ast$}}
\put(127.17,38.20){\makebox(0,0){$\ast$}}
\put(132.89,39.31){\makebox(0,0){$\ast$}}
\put(138.60,38.99){\makebox(0,0){$\ast$}}
\put(144.32,37.89){\makebox(0,0){$\ast$}}
\put(150.03,40.89){\makebox(0,0){$\ast$}}
\put(155.75,40.89){\makebox(0,0){$\ast$}}
\put(161.47,42.31){\makebox(0,0){$\ast$}}
\put(167.18,43.25){\makebox(0,0){$\ast$}}
\put(172.90,45.15){\makebox(0,0){$\ast$}}
\put(178.61,49.25){\makebox(0,0){$\ast$}}
\put(184.33,49.10){\makebox(0,0){$\ast$}}
\put(190.04,54.46){\makebox(0,0){$\ast$}}
\put(195.76,63.46){\makebox(0,0){$\ast$}}
\put(201.48,113.34){\makebox(0,0){$\ast$}}
\put(207.19,101.82){\makebox(0,0){$\ast$}}
\put(212.91,109.71){\makebox(0,0){$\ast$}}
\put(218.62,119.81){\makebox(0,0){$\ast$}}
\put(224.34,121.70){\makebox(0,0){$\ast$}}
\put(230.05,122.34){\makebox(0,0){$\ast$}}
\put(65.73,20.02){\makebox(0,0){$\star$}}
\put(71.45,20.22){\makebox(0,0){$\star$}}
\put(77.16,21.67){\makebox(0,0){$\star$}}
\put(82.88,21.33){\makebox(0,0){$\star$}}
\put(88.59,25.79){\makebox(0,0){$\star$}}
\put(94.31,26.86){\makebox(0,0){$\star$}}
\put(100.02,28.31){\makebox(0,0){$\star$}}
\put(105.74,30.45){\makebox(0,0){$\star$}}
\put(111.45,29.62){\makebox(0,0){$\star$}}
\put(117.17,30.06){\makebox(0,0){$\star$}}
\put(122.89,32.53){\makebox(0,0){$\star$}}
\put(128.60,28.85){\makebox(0,0){$\star$}}
\put(134.32,35.73){\makebox(0,0){$\star$}}
\put(140.03,34.76){\makebox(0,0){$\star$}}
\put(145.75,36.07){\makebox(0,0){$\star$}}
\put(151.46,37.82){\makebox(0,0){$\star$}}
\put(157.18,41.70){\makebox(0,0){$\star$}}
\put(162.89,43.64){\makebox(0,0){$\star$}}
\put(180.04,50.96){\makebox(0,0){$\star$}}
\put(185.76,55.95){\makebox(0,0){$\star$}}
\put(191.47,63.13){\makebox(0,0){$\star$}}
\put(197.19,68.46){\makebox(0,0){$\star$}}
\put(202.90,119.32){\makebox(0,0){$\star$}}
\put(211.48,118.88){\makebox(0,0){$\star$}}
\put(222.91,122.13){\makebox(0,0){$\star$}}
\put(94.31,26.33){\makebox(0,0){$\times$}}
\put(100.02,28.58){\makebox(0,0){$\times$}}
\put(105.74,32.19){\makebox(0,0){$\times$}}
\put(111.45,31.08){\makebox(0,0){$\times$}}
\put(117.17,33.62){\makebox(0,0){$\times$}}
\put(122.89,36.02){\makebox(0,0){$\times$}}
\put(128.60,38.47){\makebox(0,0){$\times$}}
\put(134.32,39.37){\makebox(0,0){$\times$}}
\put(140.03,39.34){\makebox(0,0){$\times$}}
\put(145.75,38.18){\makebox(0,0){$\times$}}
\put(151.46,39.71){\makebox(0,0){$\times$}}
\put(157.18,39.88){\makebox(0,0){$\times$}}
\put(162.89,42.25){\makebox(0,0){$\times$}}
\put(168.61,42.50){\makebox(0,0){$\times$}}
\put(174.33,44.73){\makebox(0,0){$\times$}}
\put(180.04,48.41){\makebox(0,0){$\times$}}
\put(185.76,48.65){\makebox(0,0){$\times$}}
\put(191.47,53.45){\makebox(0,0){$\times$}}
\put(197.19,63.01){\makebox(0,0){$\times$}}
\put(202.90,112.34){\makebox(0,0){$\times$}}
\put(208.62,101.07){\makebox(0,0){$\times$}}
\put(214.34,109.38){\makebox(0,0){$\times$}}
\put(220.05,120.10){\makebox(0,0){$\times$}}
\put(225.77,120.73){\makebox(0,0){$\times$}}
\put(231.48,121.62){\makebox(0,0){$\times$}}
\put(48.58,18.96){\makebox(0,0){$\diamond$}}
\put(54.30,20.90){\makebox(0,0){$\diamond$}}
\put(60.01,23.51){\makebox(0,0){$\diamond$}}
\put(65.73,25.11){\makebox(0,0){$\diamond$}}
\put(71.45,25.21){\makebox(0,0){$\diamond$}}
\put(77.16,27.10){\makebox(0,0){$\diamond$}}
\put(82.88,28.65){\makebox(0,0){$\diamond$}}
\put(88.59,29.72){\makebox(0,0){$\diamond$}}
\put(94.31,32.87){\makebox(0,0){$\diamond$}}
\put(100.02,34.71){\makebox(0,0){$\diamond$}}
\put(105.74,35.88){\makebox(0,0){$\diamond$}}
\put(111.45,32.63){\makebox(0,0){$\diamond$}}
\put(117.17,37.28){\makebox(0,0){$\diamond$}}
\put(122.89,39.08){\makebox(0,0){$\diamond$}}
\put(128.60,42.42){\makebox(0,0){$\diamond$}}
\put(134.32,43.15){\makebox(0,0){$\diamond$}}
\put(145.75,40.63){\makebox(0,0){$\diamond$}}
\put(151.46,45.82){\makebox(0,0){$\diamond$}}
\put(157.18,43.78){\makebox(0,0){$\diamond$}}
\put(162.89,45.33){\makebox(0,0){$\diamond$}}
\put(168.61,43.25){\makebox(0,0){$\diamond$}}
\put(174.33,48.39){\makebox(0,0){$\diamond$}}
\put(180.04,48.05){\makebox(0,0){$\diamond$}}
\put(185.76,51.20){\makebox(0,0){$\diamond$}}
\put(191.47,52.70){\makebox(0,0){$\diamond$}}
\put(197.19,54.50){\makebox(0,0){$\diamond$}}
\put(180.04,52.36){\makebox(0,0){$\triangleleft$}}
\put(185.76,54.30){\makebox(0,0){$\triangleleft$}}
\put(191.47,58.67){\makebox(0,0){$\triangleleft$}}
\put(197.19,65.45){\makebox(0,0){$\triangleleft$}}
\put(200.05,71.66){\makebox(0,0){$\triangleleft$}}
\put(208.62,97.70){\makebox(0,0){$\triangleleft$}}
\put(211.48,106.57){\makebox(0,0){$\triangleleft$}}
\put(222.91,116.36){\makebox(0,0){$\triangleleft$}}
\put(97.17,24.29){\makebox(0,0){$\triangleright$}}
\put(108.60,25.70){\makebox(0,0){$\triangleright$}}
\put(120.03,28.61){\makebox(0,0){$\triangleright$}}
\put(131.46,29.14){\makebox(0,0){$\triangleright$}}
\put(142.89,33.41){\makebox(0,0){$\triangleright$}}
\put(154.32,35.20){\makebox(0,0){$\triangleright$}}
\put(165.75,38.50){\makebox(0,0){$\triangleright$}}
\put(177.18,40.92){\makebox(0,0){$\triangleright$}}
\put(98.59,27.15){\makebox(0,0){$\odot$}}
\put(110.03,30.06){\makebox(0,0){$\odot$}}
\put(117.17,32.97){\makebox(0,0){$\odot$}}
\put(122.89,33.45){\makebox(0,0){$\odot$}}
\put(128.60,34.91){\makebox(0,0){$\odot$}}
\put(132.89,35.88){\makebox(0,0){$\odot$}}
\put(135.75,36.85){\makebox(0,0){$\odot$}}
\put(138.60,37.33){\makebox(0,0){$\odot$}}
\put(141.46,38.30){\makebox(0,0){$\odot$}}
\put(144.32,38.79){\makebox(0,0){$\odot$}}
\put(147.18,39.27){\makebox(0,0){$\odot$}}
\put(151.46,40.24){\makebox(0,0){$\odot$}}
\put(157.18,41.70){\makebox(0,0){$\odot$}}
\put(162.89,42.67){\makebox(0,0){$\odot$}}
\put(168.61,44.12){\makebox(0,0){$\odot$}}
\put(174.33,45.58){\makebox(0,0){$\odot$}}
\put(180.04,47.03){\makebox(0,0){$\odot$}}
\put(187.19,51.39){\makebox(0,0){$\odot$}}
\put(195.76,62.06){\makebox(0,0){$\odot$}}
\put(205.76,95.03){\makebox(0,0){$\odot$}}
\put(217.19,108.60){\makebox(0,0){$\odot$}}
\put(227.20,109.57){\makebox(0,0){$\odot$}}
\normalsize
\end{picture}
\end{center}
\normalsize
\caption[]{Elastic $S$-wave $\pi\pi$ phase shifts. The various
sets of data are taken from ($\odot$, \cite{PRD7p1279}),
($\star$, $\times$, $\diamond$, $\triangleleft$,
$\triangleright$, for analyses A, B, C, D, and E, respectively, of
\cite{NPB75p189}), ($\circ$, \cite{NPB79p301}),
($\ast$, \cite{NPB64p134}), and
($\bullet$, \cite{PRL47p1378}).
The solid line is our model result.}
\label{pipis}
\end{figure}
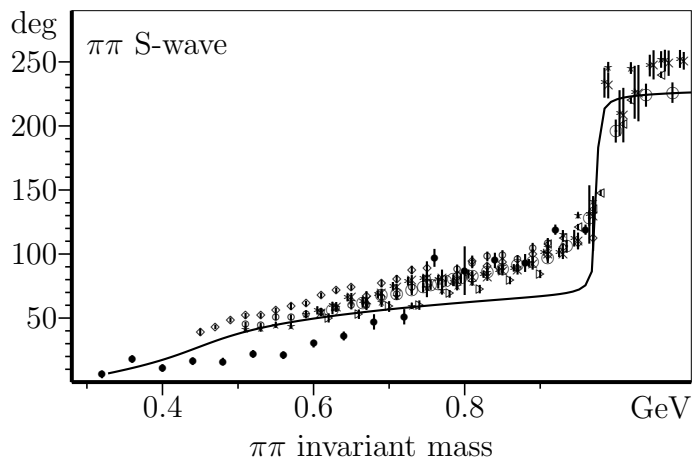

Let us first discuss $S$-wave $\pi\pi$ scattering.
The lowest bound state of our confining potential for $J^{PC}=0^{++}$
$q\bar{q}$ pairs has a mass of about 1.3 GeV,
which is at precisely the same place as the ground state of other
bound-state meson models. If we turn on the overall coupling constant
of the transition potential, bound states show up as resonances in
$\pi\pi$ scattering. At the model value of the overall coupling constant,
which was obtained from the analysis of pseudoscalar and vector mesons
\cite{PRD27p1527}, a pole shows up with a real part of about 1.3 GeV,
which accidentally equals the above-mentioned bound-state mass.
Naively we might expect that one would only find a resonating structure
in $\pi\pi$ scattering in that energy domain. However, Fig.~(\ref{pipis})
shows that the calculated phase shifts have structures
at much lower energies, which indicates that low-lying resonance poles
have been generated.

We can scan the complex energy plane for these poles
in the scattering matrix, finding one pole at about 450 MeV with a roughly
250 MeV imaginary part, and another pole at the $S$(980) position.
The imaginary part of the first pole is so large that a simple
Breit-Wigner parametrization is impossible, and large differences between
the ``mass'' of the resonance and the real part of the pole position
will occur. How these poles are connected to the harmonic-oscillator
bound states is a very technical story, which is beyond the scope of
this paper; suffice it to state that such a connection exists.
As we have discussed in \cite{LNP211p331}, these poles are special features
of $S$-wave scattering and do not show up in $P$- and higher-wave scattering,
which explains quite naturally why they are not found there.

Figure~(\ref{pipis}) also shows that the new structures at low energies
are in reasonable agreement with the experimental situation.
A criticism which might come up if one inspects this figure in more detail
is that the theoretical phase shifts do not fit the data to a
high precision, but only roughly follow the experimental slope in the data.
However, this is not a fair criticism, since we are comparing our calculations
with the {\it raw data}, with unsubtracted background,
which are presently the only available data.

In our model we left out of consideration all possible final-state
interactions in the scattering channels, like meson exchange,
Pomeron exchange, quark interchange, etc..
Moreover, the form of the transition potential may be too simple, since,
for example, we have taken a local transition potential, and it could
also have a more complicated $r$ dependence due to more sophisticated
meson-decay form factors. So our calculations better do not follow the data
very accurately. It remains, however, a pity that no analysis exists for
meson-meson scattering which subtracts the known effects and leaves us
with the consequences of the remaining interactions, a strategy that
is nowadays popular in analyzing nucleon-nucleon scattering data
\cite{LNP197p390}, because then we could really see how good the remaining
interactions are accounted for in our approach.
\mbox{From} our present calculations, we must conclude that final-state
interactions will probably alter the phase shifts a bit in the region around
600~MeV in order to change the slope of the curve towards the data.
Note that the phase shifts for low energies are almost completely
accounted for by the coupling to the permanently closed $q\bar{q}$ channels.

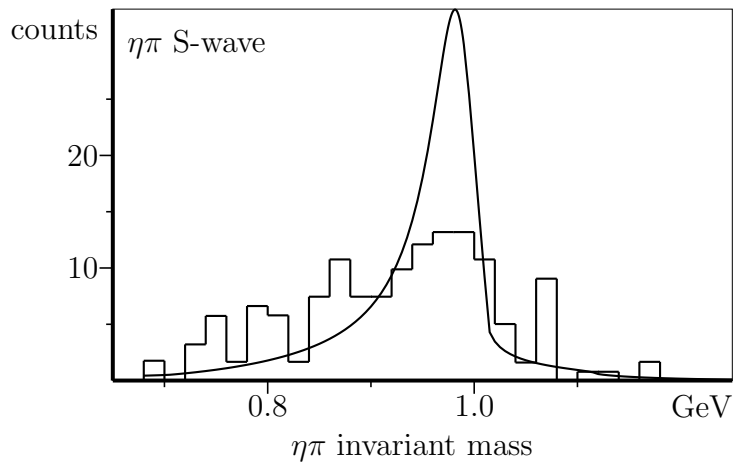
\begin{figure}[ht]
\normalsize
\begin{center}
\begin{picture}(283.46,180.08)(-50.00,-30.00)
\put(58.59,-5.52){\makebox(0,0)[tc]{0.8}}
\put(136.70,-5.52){\makebox(0,0)[tc]{1.0}}
\put(-5.52,42.61){\makebox(0,0)[rc]{10}}
\put(-5.52,85.21){\makebox(0,0)[rc]{20}}
\put(234.34,-5.52){\makebox(0,0)[tr]{GeV}}
\put(113.16,-20.07){\makebox(0,0)[tc]{$\eta\pi$ invariant mass}}
\put(-5.52,136.59){\makebox(0,0)[tr]{counts}}
\put(5.52,132.57){\makebox(0,0)[tl]{$\eta\pi$ S-wave}}
\end{picture}
\end{center}
\normalsize
\caption[]{Elastic $S$-wave $\eta\pi$ cross section. The solid line is our
model calculation. Data are taken from \cite{NPB144p253}.}
\label{etapis}
\end{figure}

In $S$-wave $\eta\pi$ scattering, the data are limited to cross sections
in the energy domain of the $\delta$. Here we find that straightforward
calculations lead to problems with the position of the $\delta$ resonance.
We suspect that these difficulties are connected to the $U(1)$ problem.
Our strategy in this case is discussed below, and
the result is depicted in Fig.~(\ref{etapis}), where we shifted the
calculated cross section by 20~MeV in order to get the peak values of
the experimental and the theoretical curves on top of each other.

\begin{figure}[ht]
\normalsize
\begin{center}
\begin{picture}(283.46,180.08)(-50.00,-30.00)
\put(0.00,-5.52){\makebox(0,0)[tc]{0.6}}
\put(36.62,-5.52){\makebox(0,0)[tc]{0.8}}
\put(73.23,-5.52){\makebox(0,0)[tc]{1.0}}
\put(109.85,-5.52){\makebox(0,0)[tc]{1.2}}
\put(146.46,-5.52){\makebox(0,0)[tc]{1.4}}
\put(183.08,-5.52){\makebox(0,0)[tc]{1.6}}
\put(-5.52,32.70){\makebox(0,0)[rc]{100}}
\put(-5.52,65.40){\makebox(0,0)[rc]{200}}
\put(-5.52,98.10){\makebox(0,0)[rc]{300}}
\put(234.34,-5.52){\makebox(0,0)[tr]{GeV}}
\put(113.16,-20.07){\makebox(0,0)[tc]{$K\pi$ invariant mass}}
\put(-5.52,136.59){\makebox(0,0)[tr]{degrees}}
\put(5.52,132.57){\makebox(0,0)[tl]{phase}}
\put(23.80,6.87){\makebox(0,0){$\odot$}}
\put(32.95,8.89){\makebox(0,0){$\odot$}}
\put(40.28,13.31){\makebox(0,0){$\odot$}}
\put(44.85,12.03){\makebox(0,0){$\odot$}}
\put(46.68,13.21){\makebox(0,0){$\odot$}}
\put(48.52,12.43){\makebox(0,0){$\odot$}}
\put(50.35,12.46){\makebox(0,0){$\odot$}}
\put(52.18,12.79){\makebox(0,0){$\odot$}}
\put(54.01,12.95){\makebox(0,0){$\odot$}}
\put(55.84,11.35){\makebox(0,0){$\odot$}}
\put(57.67,10.95){\makebox(0,0){$\odot$}}
\put(59.50,12.16){\makebox(0,0){$\odot$}}
\put(61.33,13.37){\makebox(0,0){$\odot$}}
\put(63.16,14.00){\makebox(0,0){$\odot$}}
\put(64.99,15.34){\makebox(0,0){$\odot$}}
\put(69.57,15.83){\makebox(0,0){$\odot$}}
\put(76.89,17.62){\makebox(0,0){$\odot$}}
\put(84.22,18.90){\makebox(0,0){$\odot$}}
\put(91.54,19.68){\makebox(0,0){$\odot$}}
\put(98.86,20.01){\makebox(0,0){$\odot$}}
\put(106.19,21.68){\makebox(0,0){$\odot$}}
\put(113.51,22.66){\makebox(0,0){$\odot$}}
\put(120.83,23.44){\makebox(0,0){$\odot$}}
\put(128.15,25.93){\makebox(0,0){$\odot$}}
\put(135.11,29.43){\makebox(0,0){$\odot$}}
\put(141.70,31.95){\makebox(0,0){$\odot$}}
\put(149.57,36.62){\makebox(0,0){$\odot$}}
\put(157.26,40.22){\makebox(0,0){$\odot$}}
\put(164.77,45.61){\makebox(0,0){$\odot$}}
\put(175.76,56.90){\makebox(0,0){$\odot$}}
\put(192.23,74.88){\makebox(0,0){$\odot$}}
\put(210.54,122.29){\makebox(0,0){$\odot$}}
\put(228.85,135.37){\makebox(0,0){$\odot$}}
\put(41.92,9.16){\makebox(0,0){$\bullet$}}
\put(43.39,10.79){\makebox(0,0){$\bullet$}}
\put(45.77,12.10){\makebox(0,0){$\bullet$}}
\put(46.68,11.44){\makebox(0,0){$\bullet$}}
\put(48.52,13.41){\makebox(0,0){$\bullet$}}
\put(50.90,12.75){\makebox(0,0){$\bullet$}}
\put(54.01,13.41){\makebox(0,0){$\bullet$}}
\put(56.39,11.77){\makebox(0,0){$\bullet$}}
\put(58.04,14.06){\makebox(0,0){$\bullet$}}
\put(60.23,13.41){\makebox(0,0){$\bullet$}}
\put(61.33,14.39){\makebox(0,0){$\bullet$}}
\put(63.16,14.71){\makebox(0,0){$\bullet$}}
\put(67.37,15.04){\makebox(0,0){$\bullet$}}
\put(70.49,15.37){\makebox(0,0){$\bullet$}}
\put(75.06,17.00){\makebox(0,0){$\bullet$}}
\put(79.46,17.66){\makebox(0,0){$\bullet$}}
\put(86.96,18.31){\makebox(0,0){$\bullet$}}
\put(94.10,18.97){\makebox(0,0){$\bullet$}}
\put(101.06,19.62){\makebox(0,0){$\bullet$}}
\put(108.02,20.93){\makebox(0,0){$\bullet$}}
\put(115.34,22.89){\makebox(0,0){$\bullet$}}
\put(122.66,24.52){\makebox(0,0){$\bullet$}}
\put(129.99,27.47){\makebox(0,0){$\bullet$}}
\put(133.65,27.79){\makebox(0,0){$\bullet$}}
\put(137.31,30.74){\makebox(0,0){$\bullet$}}
\put(141.52,31.72){\makebox(0,0){$\bullet$}}
\put(145.00,34.66){\makebox(0,0){$\bullet$}}
\put(148.29,37.93){\makebox(0,0){$\bullet$}}
\put(151.95,42.18){\makebox(0,0){$\bullet$}}
\put(155.62,43.82){\makebox(0,0){$\bullet$}}
\put(160.01,45.78){\makebox(0,0){$\bullet$}}
\put(162.94,49.05){\makebox(0,0){$\bullet$}}
\put(167.15,50.36){\makebox(0,0){$\bullet$}}
\put(171.18,50.03){\makebox(0,0){$\bullet$}}
\put(175.02,52.64){\makebox(0,0){$\bullet$}}
\put(178.87,54.93){\makebox(0,0){$\bullet$}}
\end{picture}
\end{center}
\normalsize
\caption[]{Elastic $I\!=\!\frac{1}{2}$ $S$-wave $K\pi$ phase shifts. Data are
from ($\odot$, \cite{PRD19p2678} and \cite{NPB133p490}) and from
($\bullet$, \cite{NPB296p493})$^{7}$.
The dashed line is our model calculation (not a fit).}
\label{Kpis}
\end{figure}
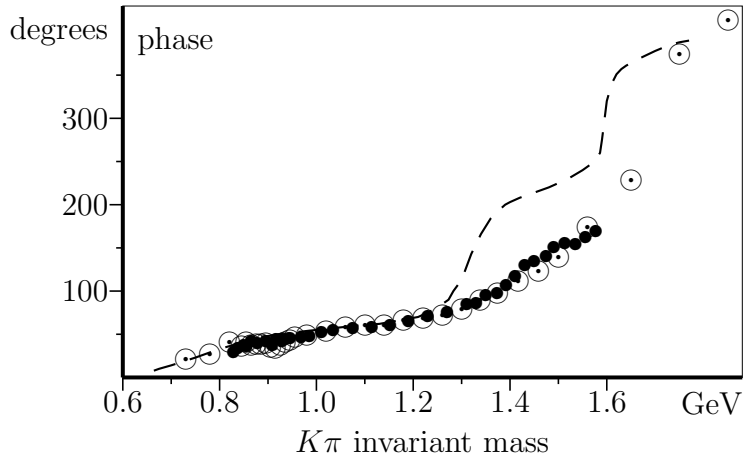
\footnotetext[7]{Added after publication.}

The results for $K\pi$ are depicted in Fig.~(\ref{Kpis}).
We see there that the phase shifts for low energies are rather well
produced by the model, but at higher energies only a rough description of
the data is given: the number of resonances at some energies agrees
with the data but the detailed structure is not reproduced at all.
Also here we have difficulties which are presumably related to the $U(1)$
problem.

In previous investigations \cite{PRD27p1527}, we took a nonstrange quark
content for the $\eta$ meson and a strange one for the $\eta'$ meson,
which is called ideal mixing. The data in the case of scalar mesons
are, however, more sensitive to the quark contents of the $\eta$'s
and we found the following: the $\eta K$ channel in the isodoublet case
is much less coupled than what would follow from our general approach as
described in \cite{PRD25p2406}, because the data show that there is not
much inelasticity below $\eta'K$. The best result is obtained if the
$\eta K$ channel is completely decoupled and $\eta'K$ enhanced to
compensate.

Something similar appears to be necessary in the isoscalar case:
if we take for $\eta$ the $s\bar{s}$ and for $\eta'$ the $n\bar{n}$ system,
we find the best results for the $S$ pole, although the complete
$(\epsilon , S)$ system is not very sensitive to these changes.
The $\delta$, however, is very sensitive to our approach, since the lowest
threshold is $\eta\pi$. In this case we have to reduce the
$\eta\pi$ coupling by a factor 1/6 and to enhance the $\eta'\pi$ coupling
to compensate. This would be the case if the $SU(3)$ mixing angle between
$\eta_{1}$ and $\eta_{8}$ was $+11^{\circ}$. The plus sign is puzzling,
because T\"{o}rnqvist \cite{PRL49p624} claimed that he needs
the standard $-11^{\circ}$ \cite{RMP56pS1} in order to fit the data.
A more rigorous study on this point is in preparation, which shows that part
of the problem might stem from our choice of transition potential.

\section{Conclusions}

If we allow only minor differences in the transition potential and take
some action with respect to the $\eta$ mesons, we find that the scattering
matrices for $S$-wave meson-meson scattering agree reasonably with the data.
An inspection of the complex energy plane shows the following poles
in the various scattering matrices: $S$($994-20i$ MeV),
$\delta$($968-28i$ MeV), $\epsilon$($470-208i$ MeV), and
$\kappa$($727-263i$ MeV).
Preliminary results show that the $\epsilon$ and $\kappa$ poles are rather
sensitive to the transition potential. The $\epsilon$ is, however, always
somewhere around 500 MeV central value, with a large width.
The $\kappa$ may vary more and even become slightly heavier than
1 GeV \cite{PRD12p14}, with a rather large width. Nevertheless, the main
conclusion is that these poles occur as normal $q\bar{q}$ configurations with
meson-meson admixtures.

Our calculations show clearly that there is no need to incorporate
$qq\bar{q}\bar{q}$ channels in our meson model. This indicates that there
is no phenomenological reason why genuine $qq\bar{q}\bar{q}$ configurations
should couple strongly to meson-meson channels.
\clearpage

\appendix
\section{Appendix}
\label{appendix}

The resonances are described by the set of Schrodinger equations

\begin{equation}
\left\{ -\fnd{d^{2}}{dr^{2}}\; +\;\fnd{L(L+1)}{r^{2}}
\; +\;2\mu (E)V(r)\; -\; k^{2}(E)\right\}\;
\phi_{E}(r)\; =\; 0
\;\;\; ,
\label{Schroedinger}
\end{equation}

\noindent
which consist of $n$ confined (permanently closed) channels and $m$ free
(open or closed scattering) channels. So $L$, $\mu$, $k^{2}$ are
$(n+m)\times (n+m)$ diagonal matrices containing the orbital angular momenta,
the reduced masses, and the momenta of the various channels.
The latter two quantities are determined from

\begin{equation}
E\; =\;\sqrt{k^{2}+{m_{1}}^{2}}\; +\;\sqrt{k^{2}+{m_{2}}^{2}}
\;\;\;\xrm{and}\;\;\;
\mu\; =\;\frac{1}{2}\fnd{dk^{2}}{dE}
\;\;\; .
\label{kinematics}
\end{equation}

Here, $m_{1}$ and $m_{2}$ stand for the masses of the quarks in the confined
channels or the meson masses in the scattering channels. The latter are
taken to be the experimental values, while the quark masses were determined
\cite{PRD27p1527}, for the $J^{PC}=0^{-+}$ pseudoscalars and $J^{PC}=1^{--}$
vectors, to be
\begin{equation}
m_{n}\; =\; 406\;\;\xrm{{MeV}} ,\;\;
m_{s}\; =\; 508\;\;\xrm{{MeV}} ,\;\;
m_{c}\; =\; 1562\;\;\xrm{{MeV}} ,\;\;\xrm{and}\;\;
m_{b}\; =\; 4724\;\;\xrm{{MeV}} \; .
\label{qmasses}
\end{equation}

The nonrelativistic limit of (\ref{kinematics}) is used in the confined
channels, and for energies lower than the threshold also in the free channels.

The $(n+m)\times (n+m)$ potential matrix reads

\begin{equation}
V\; =\;\left(\begin{array}{cc}
V_{c} & V_\xrm{{\tiny int}} \\ [.3cm]
V_\xrm{{\tiny int}}^{T} & V_{f}\end{array}\right)
\;\;\; .
\label{potential}
\end{equation}

The confining potential $V_{c}$ is a diagonal $n\times n$ matrix containing
the mass-dependent harmonic oscillators

\begin{equation}
\left[ V_{c}\right]_{ii}\; =\; \frac{1}{2}\mu_{i}\omega^{2}r^{2}
\;\;\; .
\label{Vc}
\end{equation}

The $m\times m$ matrix $V_{f}$ describes possible final-state interactions.
The $n\times m$ transition potential $V_\xrm{{\tiny int}}$ is taken to be

\begin{equation}
\left[ V_\xrm{{\tiny int}}\right]_{ij}\; =\;
\tilde{g}\omega\; c_{ij}\;\sqrt{\fnd{E}{D}}\;\fnd{r}{r_{0}}\;
\exp\left\{ -\frac{1}{2}\left(\fnd{r}{r_{0}}\right)^{2}\right\}
\;\;\; ,
\label{Vint}
\end{equation}

\noindent
where $\tilde{g}$ is the universal coupling constant,
$r_{0}=\rho_{0}/\sqrt{\mu_{i}\omega}$ the transition radius, and
$E/D$ a phenomenological factor, with $D=m_{1}+m_{2}$. The above-mentioned
parameters were also fixed \cite{{PRD27p1527}} by the $J^{PC}=0^{-+}$
pseudoscalars and $J^{PC}=1^{--}$ vectors to

\begin{equation}
\omega\; =\; 190\;\;\xrm{{MeV}} ,\;\;
\tilde{g}\; =\; 10.4 ,\;\;\xrm{and}\;\;
\rho_{0}\; =\; 0.56\;\; .
\label{param}
\end{equation}

In comparison to the $0^{-+}$ and $J^{PC}=1^{--}$ case, we left out the
color term, which was necessary for the pseudoscalar/vector splitting.
This is, of course, not a very serious intervention, since for $P$-wave
$q\bar{q}$ systems the color splitting is anyhow much smaller than for
$S$-wave systems \cite{PRD12p14}.

The numbers $c_{ij}$ are the relative decay couplings. If we pursue the
concepts behind the confining potential further to the phase of
the pair creation, we are led to the assumption that all the interquark
forces are harmonic-oscillator forces during this process.
This idea provides us a scheme in which the possible decay channels
and their relative strengths can be calculated \cite{PRD25p2406}.
Their values are listed in Table~\ref{couplings}.

\begin{table}[ht]
\begin{center}
\begin{tabular}{|l|l||l|l|l|l|}
\hline
\multicolumn{2}{|c||}{Decay Products} &
\multicolumn{4}{c|}{Initial Mesons} \\ [.3cm]
Channel & Spin & \multicolumn{2}{c|}{$\epsilon$ and $S$} &
$\delta$ & $\kappa$ \\ [.3cm]
& & $n\bar{n}$ & $s\bar{s}$ & & \\ \hline & & & & & \\
$\pi\pi$ & 0 & 3/40 & & & \\ [.1cm]
$\eta_{n}\eta_{n}$ & 0 & 1/40 & & & \\ [.1cm]
$\eta_{s}\eta_{s}$ & 0 & & 1/16 & & \\ [.1cm]
$KK$ & 0 & 1/40 & 1/16 & 1/24 & \\ [.1cm]
$\rho\rho$ & 0 & 1/40 & & & \\ [.1cm]
$\rho\rho$ & 2 & 1/2 & & & \\ [.1cm]
$\omega\omega$ & 0 & 1/120 & & & \\ [.1cm]
$\omega\omega$ & 2 & 1/6 & & & \\ [.1cm]
$K^{\ast}K^{\ast}$ & 0 & 1/120 & 1/48 & 1/72 & \\ [.1cm]
$K^{\ast}K^{\ast}$ & 2 & 1/6 & 5/12 & 5/18 & \\ [.1cm]
$\phi\phi$ & 0 & & 1/48 & & \\ [.1cm]
$\phi\phi$ & 2 & & 5/12 & & \\ [.1cm]
$\eta_{n}\pi$ & 0 & & & 1/12 & \\ [.1cm]
$\eta_{s}\pi$ & 0 & & & & \\ [.1cm]
$\rho\omega$ & 0 & & & 1/36 & \\ [.1cm]
$\rho\omega$ & 2 & & & 5/9 & \\ [.1cm]
$\pi K$ & 0 & & & & 1/16 \\ [.1cm]
$\eta_{n}K$ & 0 & & & & 1/48 \\ [.1cm]
$\eta_{s}K$ & 0 & & & & 1/24 \\ [.1cm]
$\rho K^{\ast}$ & 0 & & & & 1/48 \\ [.1cm]
$\rho K^{\ast}$ & 2 & & & & 5/12 \\ [.1cm]
$\omega K^{\ast}$ & 0 & & & & 1/144 \\ [.1cm]
$\omega K^{\ast}$ & 2 & & & & 5/36 \\ [.1cm]
$\phi K^{\ast}$ & 0 & & & & 1/72 \\ [.1cm]
$\phi K^{\ast}$ & 2 & & & & 5/18 \\
& & & & & \\ \hline
\end{tabular}
\end{center}
\caption[]{Relative quadratic coupling constants $c_{ij}$ (see text)
for the decay process of a scalar
meson into a pair of pseudoscalar/vector mesons.}
\label{couplings}
\end{table}

The $\eta_{n}$ and $\eta_{s}$ denote the pure $n\bar{n}$ and $s\bar{s}$
states, respectively. Of course, these are not the physical $\eta$ and
$\eta'$. See the Conclusions for a discussion about this point.
The $\pi\pi$ ($S$-wave) scattering is described in one system of equations,
with two confined channels containing an $n\bar{n}$ and an $s\bar{s}$ pair.
These systems are linked to each other via decay in which strange mesons
occur.
\clearpage

\def\LNC{Lett.\ Nuovo Cim.}
\def\SJNP{Sov.\ J.\ Nucl.\ Phys.}
\def\ZPC{Z.\ Phys.\ C}

\newcommand{\pubprt}[4]{#1 {\bf #2}, #3 (#4)}
\newcommand{\ertbid}[4]{[Erratum-ibid.\ #1 {\bf #2}, #3 (#4)]}
\def\CPC{Comput.\ Phys.\ Commun.}
\def\LNP{Lect.\ Notes Phys.}
\def\NPB{Nucl.\ Phys.\ B}
\def\PLB{Phys.\ Lett.\ B}
\def\PRD{Phys.\ Rev.\ D}
\def\PRL{Phys.\ Rev.\ Lett.}
\def\RMP{Rev.\ Mod.\ Phys.}


\begin{thebibliography}{32}
\bibitem{NPB121p514}
A.~D.~Martin, E.~N.~Ozmutlu and E.~J.~Squires,
\pubprt{\NPB}{121}{514}{1977};
N.~N.\ Achasov, S.~A.~Devyanin and G.~N.~Shestakov,
\pubprt{\PLB}{88}{367}{1979};
\pubprt{\SJNP}{32}{566}{1980}
[Yad.\ Fiz.\  {\bf 32}, 1098 (1980)].

\bibitem{PRL49p624}
N.~A.~T\"{o}rnqvist,
\pubprt{\PRL}{49}{624}{1982};
K.~Heikkil\"{a}, S.~Ono and N.~A.~T\"{o}rnqvist,
\pubprt{\PRD}{29}{110}{1984}
\ertbid{\ D}{29}{2136}{1984}.

\bibitem{PLB51p71}
D.~Morgan,
\pubprt{\PLB}{51}{71}{1974}.

\bibitem{PRD19p2678}
P.~Estabrooks,
\pubprt{\PRD}{19}{2678}{1979}.

\bibitem{PLB93p65}
A.~Bramon and E.~Masso,
\pubprt{\PLB}{93}{65}{1980}
\ertbid{\ B}{107}{455}{1981};
\pubprt{\PLB}{120}{240}{1983}.

\bibitem{PRD15p267}
R.~L.~Jaffe,
\pubprt{\PRD}{15}{267}{1977};
281 (1977);
{\bf17}, 1444 (1978).

\bibitem{PLB60p201}
R.~L.~Jaffe and K.~Johnson,
\pubprt{\PLB}{60}{201}{1976}.

\bibitem{PLB50p1}
V.~Chaloupka {\it et al.} \/[Particle Data Group],
\pubprt{\PLB}{50}{1}{1974}.

\bibitem{PRD7p1279}
S.~D.~Protopopescu {\it et al.},
\pubprt{\PRD}{7}{1279}{1973}.

\bibitem{NPB75p189}
G.~Grayer {\it et al.},
\pubprt{\NPB}{75}{189}{1974}.

\bibitem{NPB79p301}
P.~Estabrooks and A.~D.~Martin,
\pubprt{\NPB}{79}{301}{1974}.

\bibitem{NPB64p134}
B.~Hyams {\it et al.},
\pubprt{\NPB}{64}{134}{1973}
[AIP Conf.\ Proc.\  {\bf 13}, 206 (1973)];
\pubprt{\NPB}{100}{205}{1975}.

\bibitem{PRL47p1378}
N.~N.~Biswas {\it et al.},
\pubprt{\PRL}{47}{1378}{1981}.

\bibitem{NPB95p322}
P.~Estabrooks and A.~D.~Martin,
\pubprt{\NPB}{95}{322}{1975};
C.~D.~Froggatt and J.~L.~Petersen,
\pubprt{\NPB}{129}{89}{1977};
M.~J.~Corden {\it et al.},
\pubprt{\NPB}{157}{250}{1979}.

\bibitem{NPB144p253}
M.~J.~Corden {\it et al.},
\pubprt{\NPB}{144}{253}{1978}.

\bibitem{NPB101p333}
J.~Wells, D.~Radojicic, D.~A.~Roscoe and L.~Lyons,
\pubprt{\NPB}{101}{333}{1975};
H.~Gr\"{a}ssler {\it et al.}
[Aachen-Berlin-Bonn-CERN-Cracow-Heidelberg-Warsaw Collaboration],
\pubprt{\NPB}{121}{189}{1977};
J.~B.~Gay {\it et al.}  [Amsterdam-CERN-Nijmegen-Oxford Collaboration],
\pubprt{\PLB}{63}{220}{1976};
A.~C.~Irving,
\pubprt{\PLB}{70}{217}{1977};
A.~Gurtu {\it et al.} [Amsterdam-CERN-Nijmegen-Oxford Collaboration],
\pubprt{\NPB}{151}{181}{1979};
C.~Evangelista {\it et al.},
\pubprt{\NPB}{178}{197}{1981}
\ertbid{\ B}{186}{594}{1981}].

\bibitem{RMP56pS1}
C.~G.~Wohl {\it et al.} [Particle Data Group],
\pubprt{\RMP}{56}{S1}{1984}.

\bibitem{NPB133p490}
P.~Estabrooks, R.~K.~Carnegie, A.~D.~Martin, W.~M.~Dunwoodie,
T.~A.~Lasinski and D.~W.~Leith,
\pubprt{\NPB}{133}{490}{1978}.

\bibitem{PRD9p1872}
M.~J.~Matison, A.~Barbaro-Galtieri, M.~Alston-Garnjost, S.~M.~Flatt\'{e},
J.~H.~Friedman, G.~R.~Lynch, M.~S.~Rabin and F.~T.~Solmitz,
\pubprt{\PRD}{9}{1872}{1974};
S.~L.~Baker {\it et al.},
\pubprt{\NPB}{99}{211}{1975}.

\bibitem{PRD21p1370}
A.~T.~Aerts, P.~J.~Mulders and J.~J.~De Swart,
\pubprt{\PRD}{21}{1370}{1980}.

\bibitem{PRL40p1543}
P.~J.~G.~Mulders, A.~T.~M.~Aerts and J.~J.~de Swart,
\pubprt{\PRL}{40}{1543}{1978};
\pubprt{\PRD}{21}{2653}{1980};
N.~N.~Achasov, S.~A.~Devyanin and G.~N.~Shestakov,
\pubprt{\PLB}{96}{168}{1980};
\pubprt{\ZPC}{16}{55}{1982};
V.~A.~Matveev and P.~Sorba,
\pubprt{\LNC}{20}{435}{1977};
H.~M.~Chan and H.~Hogaasen,
\pubprt{\NPB}{136}{401}{1978};
H.~Hogaasen and P.~Sorba,
\pubprt{\NPB}{145}{119}{1978}.

\bibitem{PhDAAerts}
A.~T.~M.~Aerts,
{\it The MIT bag model and some spectroscopic applications},
Ph.D. thesis, University of Nijmegen (1979);
J.~S.~Kang and H.~J.~Schnitzer,
\pubprt{\PRD}{12}{841}{1975};
D.~P.~Stanley and D.~Robson,
\pubprt{\PRD}{21}{3180}{1980}.

\bibitem{PRD27p1527}
E.~van Beveren, G.~Rupp, T.~A.~Rij\-ken and C.~Dullemond,
\pubprt{\PRD}{27}{1527}{1983}.

\bibitem{PRD21p772}
E.~van Beveren, C.~Dullemond and G.~Rupp,
\pubprt{\PRD}{21}{772}{1980}
\ertbid{\ D}{22}{787}{1980};
G.~Rupp,
{\it Spectra and decay properties of pseudoscalar and vector
mesons in a multichannel quark model},
Ph.D. thesis, University of Nijmegen (1982);
E.~van~Beveren,
{\it On the influence of hadronic decay on the properties of hadrons
(A study in the context of a geometrical quark model)},
Ph.D. thesis, University of Nijmegen (1983);
C.~Dullemond, T.~A.~Rij\-ken, E.~van Beveren and G.~Rupp,
in Proc.\ {\it VIth Warsaw Symposium on Elementary Particle Physics,}
Kazimierz, Poland, 30 May -- 3 June 1983, pp.\ 257--262;
E.~van Beveren, C.~Dullemond and T.~A.~Rij\-ken,
\pubprt{\ZPC}{19}{275}{1983}.

\bibitem{CPC27p377}
C.~Dullemond, G.~Rupp, T.~A.~Rij\-ken and E.~van Beveren,
\pubprt{\CPC}{27}{377}{1982}.

\bibitem{LNP211p182}
E.~van Beveren, C.~Dullemond, T.~A.~Rij\-ken and G.~Rupp,
\pubprt{\LNP}{211}{182}{1984}.

\bibitem{LNP211p331}
Reference \cite{LNP211p182}, p.\ 331.

\bibitem{PRD25p2406}
J.~E.~Ribeiro,
\pubprt{\PRD}{25}{2406}{1982};
E.~van Beveren,
\pubprt{\ZPC}{17}{135}{1983}
[arXiv:hep-ph/0602248];
\pubprt{\ZPC}{21}{291}{1984}
[arXiv:hep-ph/0602247].

\bibitem{NPB10p521}
L.~Micu,
\pubprt{\NPB}{10}{521}{1969};
R.~D.~Carlitz and M.~Kislinger,
\pubprt{\PRD}{2}{336}{1970}.

\bibitem{LNP197p390}
J.~R.~Bergervoet  {\it et al.},
in Proc. {\it Conference on Quarks and Nuclear Structure},
Bad Honnef (1983, Germany),
\pubprt{\LNP}{197}{390}{1984}.

\bibitem{NPB296p493}
D.~Aston {\it et al.} [LASS collaboration],
\pubprt{\NPB}{296}{493}{1988}.

\bibitem{PRD12p14}
A.~de R\'{u}jula, H.~Georgi and S.~L.~Glashow
\pubprt{\PRD}{12}{14}{1975}.
\end{thebibliography}
\end{document}